\documentstyle[12pt]{article}

\begin{document}
\newcommand{\sh}{\mbox{sinh} \tau  }
\newcommand{\ch}{\mbox{cosh} \tau  }
\newcommand{\si}{\mbox{sin} \sigma  }
\newcommand{\co}{\mbox{cos} \sigma }
\newcommand{\Sh}{\mbox{sinh} \gamma }
\newcommand{\Ch}{\mbox{cosh} \gamma }
\newcommand{\Si}{\mbox{sin} \theta  }
\newcommand{\Co}{\mbox{cos} \theta }

\newcommand{\tr}{\mbox{tr}}

\newcommand{\g}{{\bf g}}

\newcommand{\ar}{a\rangle}
\newcommand{\br}{b\rangle}
\newcommand{\xr}{x\rangle}
\newcommand{\yr}{y\rangle}
\newcommand{\xl}{\langle x}
\newcommand{\yl}{\langle y}

\newcommand{\car}{a^*\rangle}
\newcommand{\cbr}{b^*\rangle}
\newcommand{\cxr}{x^*\rangle}
\newcommand{\cyr}{y^*\rangle}
\newcommand{\cxl}{\langle x^*}
\newcommand{\cyl}{\langle y^*}
\begin{flushright}
{LPTHE 93-44}
\end{flushright}
\begin{center}
{\bf MULTI-STRING SOLUTIONS BY SOLITON METHODS IN DE SITTER SPACETIME}
{~~~~~~}\\
\end{center}
\begin{center}
{{\bf  F. Combes$^{(a)}$,  H.J. de Vega$^{(b)}$,
  A. V. Mikhailov$^{(a,c)}$ and N. S\'anchez$^{(a)}$}}\\
{~~~~~}
\end{center}
\begin{center}

{\it (a) Observatoire de Paris, Section de Meudon, Demirm,
Laboratoire Associ\'e au CNRS UA 336, Observatoire de Meudon et
\'Ecole Normale Sup\'erieure.
 92195 MEUDON Principal Cedex, FRANCE.}

{~~~~~}

{\it (b)  Laboratoire de Physique Th\'eorique et Hautes Energies,
Universit\'e Paris VI, Tour 16, 1er \'etage, 4, Place Jussieu
75252 Paris, Cedex 05, FRANCE. }

{~~~~~}

{\it (c) Landau Institute for Theoretical Physics,  Russian Academy of
Sciences,  Ul. Kossyguina 2, 117334 Moscow, RUSSIA}\\

\end{center}
\begin{abstract}
{\bf Exact} solutions of the string equations of motion and constraints
are {\bf systematically} constructed in de Sitter spacetime using
the dressing method of soliton theory. The string dynamics in
de Sitter spacetime is integrable due to the associated linear system.
We start from an exact string solution $q_{(0)}$ and the associated
solution of the linear system $\Psi^{(0)}(\lambda)$, and we construct
a new solution $\Psi(\lambda)$ differing from   $\Psi^{(0)}(\lambda)$
by a rational matrix in $\lambda$ with at least four poles
$ \lambda _{0}, 1/\lambda _{0}, \lambda _{0}^* , 1/\lambda _{0} ^* $. The
periodicity condition for closed strings restrict $ \lambda _{0}$
to discrete values expressed in terms of
Pythagorean numbers. Here we explicitly construct
solutions depending on $(2+1)$-spacetime coordinates,
two arbitrary complex numbers
(the 'polarization vector') and two integers $(n,m)$ which determine the
string windings in the space. The solutions are depicted in the
hyperboloid coordinates $q$ and in comoving coordinates with the
cosmic time $T$ . Despite of the fact that we have a single world sheet,
our solutions describe {\bf multiple} (here five) different and
independent strings; the world sheet time $\tau$ turns to be a multivalued
function of $T$. (This has no analogue in flat spacetime).
One string is stable (its proper size tends to a constant for
$T \to \infty$, and its comoving size contracts); the other strings
are unstable (their proper sizes blow up for $T \to \infty$,
while their comoving sizes tend to constants). These solutions
(even the stable strings) do not oscillate in time.
The interpretation of these solutions
and their dynamics in terms of the sinh-Gordon model is
 particularly enlighting.
 \end{abstract}

\newpage

\section{Introduction}
\

Since the propagation of strings in curved spacetimes
started to be systematically investigated, a variety of new physical
phenomena appeared \cite{dvs87}-\cite{eri}.  These results  are relevant
both for fundamental (quantum) strings and for cosmic strings which
behave essentially in a classical way.

String propagation has been investigated in non-linear gravitational
plane waves \cite{opg} and shock-waves
\cite{sw}, black holes \cite{agn},  conical spacetimes
\cite{con},   and cosmological spacetimes
\cite{dvs87},\cite{gsv}.

Among the cosmological backgrounds, de Sitter
spacetime occupies a special
place. This is, in one hand  relevant for inflation and on the
other hand string propagation turns to be specially
interesting  there \cite{dvs87},\cite{gsv}.
String unstability, in the sense that the string proper
length grows indefinitely is particularly present in de Sitter.
The string dynamics in de Sitter universe is described by a
generalized sinh-Gordon model with a potential  unbounded from
below \cite{prd}. The sinh-Gordon function $\alpha(\sigma,\tau)$
having a clear physical meaning : $ H^{-1} e^{\alpha(\sigma,\tau)/2} $
determines the string proper length.
Moreover the classical string equations of motion (plus the
string constraints) turn to be integrable in de Sitter universe
\cite{prd},\cite{dms}. More precisely, they are equivalent to a
non-linear sigma model on the grassmannian $SO(D,1)/O(D)$ with periodic
boundary conditions (for closed strings). This sigma model
has an associated linear system \cite{zm1} and using it, one can show the
presence  of an infinite number of conserved quantities \cite{dv78}.
In addition, the string constraints imply
a zero energy-momentum tensor and these constraints
are compatible with the integrability.

The so-called dressing method \cite{zm1} in soliton theory
allows to construct solutions of non-linear classically
integrable models using the associated linear
system. In the present paper we systematically construct string solutions
in three dimensional de Sitter spacetime.
We start from a given exactly known solution
of the string equations of motion and constraints in de Sitter \cite{dms}
and then we ``dress'' it.
The string solutions reported here indeed apply to cosmic strings
in de Sitter spacetime as well. The dynamics of cosmic strings
in expanding universes has been studied in the literature for the
Friedman-Robertson-Walker (FRW)  cases (see for example \cite{twbk},
\cite{vil},\cite{ani}).
It must be noticed that the string behaviour we found
here in de Sitter universe is essentially {\bf different}
from the standard FRW where the expansion factor $R(T)$
is a positive power of the cosmic
time $T$. In such FRW universes, strings always oscillate in time,
 the comoving spatial string coordinates {\it contract} and the proper
string size stays {\it constant} asymptotically for  $T \to \infty$
\cite{gsv},\cite{ani} .
In the cosmic string literature this is known as 'string stretching'.
We called such behaviour 'stable' \cite{gsv},\cite{dms},\cite{ani}.
 On the contrary, in
de Sitter spacetime, as we show below, two
types of asymptotic behaviors are present : (i) the proper string size
and energy grow with the expansion factor ('unstable' behaviour) or
(ii) they tend to constant values ('stable' strings).

The unstable string solutions in de Sitter universe may provide a
mechanism to self-sustain inflation as proposed in
refs.\cite{tuba}-\cite{gsv}  without advocating an inflaton field. The
multi-string exact solutions in de Sitter spacetime presented here
should provide essential clues about the feasability of inflationary
string scenarios.

We apply here the dressing method as follows. We start from the exact
ring-shaped string solution $ q_{(0)} $ \cite{dms} and we find the
explicit solution $ \Psi^{(0)}(\lambda) $ of the
associated linear system, where $\lambda$ stands for the spectral
parameter.
Then, we propose a new solution $ \Psi(\lambda) $ that differs from
$ \Psi^{(0)}(\lambda) $ by a matrix rational in $\lambda$.
Notice that  $ \Psi(\lambda = 0) $ provides
in general a new string solution.
We then show that this rational matrix
must have at least {\bf four} poles,
 $ \lambda _{0}, 1/\lambda _{0}, \lambda _{0}^* , 1/\lambda _{0} ^* $,
as a consequence of the symmetries of the problem.
The residues of these poles are shown to be one-dimensional
projectors. We then prove that these projectors are formed by vectors
which can all be expressed in terms of an arbitrary complex constant
vector $ |x_0 \rangle $ and the complex parameter $ \lambda _{0} $.
This result holds for arbitrary starting solutions $ q_{(0)} $.

Since we consider closed strings, we impose a $2\pi$-periodicity on
the string variable $ \sigma $ . This restricts $ \lambda _{0} $ to
take discrete values that we succeed to express in terms
of  Pythagorean numbers.
In summary, our solutions depend on two arbitrary complex numbers
contained in  $ |x_0 \rangle $ and two integers $ n $ and $ m $ . The
counting of degrees of freedom is analogous to $2 + 1$ Minkowski
spacetime except that left and right modes are here mixed up  in a
non-linear and precise way.

The vector $ |x_0 \rangle $ somehow indicates the polarization of the
string. The integers $(n,m)$  determine the string winding. They fix
the way in which the string winds around the origin in the spatial
dimensions (here $S^2$ ). Our starting solution $q_{(0)}(\sigma,
\tau)$ is a stable string
winded $n^2 + m^2$ times  around the origin in de Sitter space.

The matrix multiplications involved in the computation of
the final solution were done with the help of  the computer
program of symbolic calculation ``Mathematica''. The resulting solution
$ q(\sigma, \tau) = (q^0, q^1, q^2, q^3) $ is a complicated
combination of trigonometric functions of $ \sigma $ and hyperbolic
functions of $\tau$. That is, these string solitonic solutions
do not oscillate in time. This is a typical feature of string
unstability \cite{agn}-\cite{gsv}-\cite{dms}. The new feature here is
that strings (even stable solutions) do not oscillate neither
for $ \tau \to 0 $, nor for $\tau \to \pm\infty $. Figs. 3-4
depict spatial projections $ (q^1, q^2, q^3) $ of the solutions
for two given polarizations $|x_0 \rangle $ and different windings
$(m, n)$.

We plot in figs. 5-11 the solutions for significative values
of  $|x_0 \rangle $ and $(m, n)$ in terms of the comoving coordinates
$( T, X^1, X^2 )$
\begin{equation}
T = \frac{1}{H}\log(q^0 + q^1) ~~,~~~ X^1 =\frac{1}{H}
\frac{q^2}{q^0 + q^1} ~~,~~X^2 =\frac{1}{H}
\frac{q^3}{q^0 + q^1} ~~
\label{cooXT}
\end{equation}
The first feature to point out is that our solitonic solutions
describe {\bf multiple} (here five  or three) strings,
as it can be seen from the fact that for a given time $T$
 we find several different values for $\tau$.
That is, $\tau$ is a {\bf multivalued} function of $T$
for any fixed $\sigma$ (fig.5-6). Each branch of $\tau$
as a function of $T$  corresponds to a different string.
This is a entirely new feature for strings in curved spacetime, with
no analogue in flat spacetime where the time coordinate
can always be chosen proportional to $\tau$.
In flat spacetime, multiple string solutions are described
by multiple world-sheets.
Here, we have a {\bf single} world-sheet describing
several independent and simultaneous strings as a consequence of the
coupling with the spacetime geometry.
Notice that we consider {\it free} strings. (Interactions
among the strings as splitting or merging  are not considered).
Five  is the generic number of strings in our dressed
solutions. The value five can be related to the fact
that we are dressing a one-string solution ($ q_{(0)} $) with {\it four}
poles. Each pole adds here an unstable string.

In order to describe the real physical evolution, we eliminated
numerically $ \tau = \tau (\sigma, T) $ from the solution and expressed
the spatial comoving coordinates $X^1$ and $X^2$ in terms of
$T$ and $\sigma$.

We plot $\tau(\sigma, T)$ as a function of $\sigma$ for different fixed values
of $T$ in fig.7-8. It is a sinusoidal-type function. Besides the
customary closed string period $2\pi$, another period appears
which varies on $\tau$. For small $\tau $ , $\tau = \tau(\sigma, T)$ has
a convoluted shape while for larger $ \tau $ (here $  \tau \leq 5
$), it becomes a regular sinusoid. These behaviours reflect very
clearly in the evolution of the spatial coordinates and shape
of the string.

The evolution of the five (and three) strings
simultaneously described by our solution as a function
of $T$, for positive $T$
 is shown in figs. 9-11. One string is stable (the 5th one). The other four are
unstable. For the stable string, $(X^1, X^2)$ contracts in time
precisely as $ e^{-HT} $, thus keeping the proper amplitude
$(e^{HT} X^1, e^{HT} X^2)$ and proper size constant.
For this stable string $(X^1, X^2) \leq {1 \over H}$.
($1/H$ = the horizon radius).  For the other
(unstable) strings, $(X^1, X^2)$ become very fast constant in time,
the proper size expanding as the universe itself like  $ e^{HT} $ .
For these strings $(X^1, X^2) \geq {1 \over H}$.
These exact solutions display remarkably the asymptotic string
behaviour found in refs.\cite{gsv},\cite{prd}.

In terms of the sinh-Gordon description, this means that for the
strings outside the horizon the sinh-Gordon function
$\alpha(\sigma,\tau)$
is the same as the cosmic time $T$ up to
a function of $\sigma$. More precisely,
\begin{equation}
\alpha(\sigma,\tau) \buildrel{T >> {1\over H} }\over =
2 H\, T(\sigma, \tau) + \log\left\{2 H^2 \left[ (A^{1}(\sigma)')^2 +
  (A^{2}(\sigma)')^2 \right] \right\} + O(e^{-2HT}).
\label{Tcosa}
\end{equation}
Here $A^1(\sigma)$ and  $A^2(\sigma)$ are the $X^1$ and $X^2$
coordinates outside the horizon. For $T \to +\infty$ these
strings are at the absolute {\it minimun} $\alpha = + \infty $
of the sinh-Gordon potential with infinite size.
The string inside the horizon (stable string) corresponds to the
 {\it maximun} of the potential, $\alpha = 0$.
$ \alpha = 0$ is the only value in which the string
can stay without being
pushed down by the potential to $\alpha = \pm \infty$ and
this also explains why only one stable
string appears (is not possible to put more than one string at
the maximun of the potential without falling down).
These features are {\it generically} exhibited by our
one-soliton multistring solutions, independently of the
particular initial state of the string
(fixed by $|x^0> $ and $(n,m)$).
For particular values of $|x^0> $, the solution describes
three strings, with symmetric shapes from $ T = 0 $, for instance
like a rosette or a circle with festoons (fig. 9-11).

The string solutions presented here trivially embedd on
D-dimensional de Sitter spacetime ($D \geq 3$). It must be noticed
that they exhibit the essential physics of strings in D-dimensional
de Sitter universe. Moreover, the construction method used here
works in any number of dimensions.

This paper is organised as follows: in section 2
we describe the string equations in de Sitter universe and its
associated linear system. Section 3 deals with the dressing
method in soliton theory, its application to this string problem
and the systematic construction of solutions. In section 4 we
explicitly describe the starting background solution
$ q_{(0)}(\sigma ,\tau) $ and the solution $ \Psi^{(0)}(\lambda) $
of the associated linear system. In section 5, we analyze our
multistring solutions and describe their physical properties.

\section{The string equations and their associated linear system}
\

The string equations of motion in  $D$-dimensional de
Sitter space-time can be written in the following form:

\begin{equation}
q_{\xi \eta }\rangle + q\rangle \langle
q_{\xi }Jq_{\eta }\rangle =0,
\label{seq}\end{equation}

\noindent where $q\rangle $ is a ($D+1$)-dimensional
real vector of unit pseudolength
\begin{equation}
 \langle qJq\rangle =1, \ \ \ (J=\mbox{diag}(-1,1,...,1))
\label{l1}\end{equation}
and $\xi , \eta$ are  light cone coordinates in the world sheet:
\begin{equation}
\tau = \eta +\xi ,\ \ \sigma = \eta - \xi
\label{lc}\end{equation}

In addition, we have the string constraints (conformal conditions)
\begin{equation}
\langle q_{\xi }Jq_{\xi }\rangle=0,\ \ \
\langle q_{\eta }Jq_{\eta }\rangle =0.
\label{scond}\end{equation}
The solution $q\rangle $ should be a periodic function of
$\sigma =   \eta - \xi$, with period $2 \pi$ for closed strings.

We are going to find solutions of this equation by using
the Riemann Transform Method \cite{zm1,zm2}. The most important
observation is that equation (\ref{seq}) can be rewritten in
the form of a chiral field model on the Grassmanian
$G_{D}=SO(D,1)/O(D)$.Indeed, any element  ${\bf g}\in G_{D}$ can be
parametrized with a real vector $q\rangle $ of the unit pseudolength
\begin{equation}
{\bf g}=1-2 q\rangle\langle qJ,\ \\\\\ \  \langle qJq\rangle =1.
\label{G}
\end{equation}
In terms of $\bf g$, the string equations (\ref{seq})-(\ref{scond})
 have the following form
\begin{equation}
2\,{\bf g}_{\xi \eta }+{\bf g}_{\xi }\,{\bf g\, g}_{\eta }+
{\bf g}_{\eta }\,{\bf g\, g}_{\xi }=0~,
\label{geq}
\end{equation}
and  the conformal constraints are
\begin{equation}
\tr \, {\bf g}_{\xi }^{2}=0,\ \ \ \tr \, {\bf g}_{\eta }^{2}=0~,
\label{gcond}\end{equation}
which are equivalent to eqs.(\ref{scond}).
The fact that ${\bf g}\in G_{D}$ implies that $\bf g $ is a
real matrix with the following properties:

\begin{equation}
{\bf g}=J\g ^{\mbox{t}}J,\ \ \ \g ^{2}=I,\ \ \ \
tr \, \g =2\ \ \ \g \in SL(D+1, R).
\label{gcond1}\end{equation}
These conditions are equivalent to the existence of the  representation
(\ref{G}).

Equation (\ref{geq}) is the compatibility condition for
the following overdetermined linear system:
\begin{equation}
\Psi _{\xi}=\frac{U}{1-\lambda }\Psi ,\ \
\Psi _{\eta}=\frac{V}{1+\lambda }\Psi ,
\label{psieq}
\end{equation}
where
\begin{equation}
U={\bf g}_{\xi}\, {\bf g},\ \ \ V={\bf g}_{\eta}\, {\bf g} \ \ .
\label{UV}\end{equation}
Or in terms of vector $q\rangle $
$$
U=2\,  q_{\xi }\rangle\langle q \,
J -2 \, q\rangle \langle q_{\xi }\, J , \ \
$$
$$
V=2\,  q_{\eta }\rangle\langle q \, J-2 \,
q\rangle \langle q_{\eta }\, J .
$$
In order to fix the freedom in the definition
of $\Psi $ we shall identify
\begin{equation}
\Psi (\lambda =0)=\bf g.
\label{l0g}\end{equation}

This condition  is compatible with the above equations since the matrix
function $\Psi $ at the point $\lambda =0$ satisfies
the same equations as $\bf g$. Thus the problem of
constructing exact solutions of the string equations
is reduced to finding  compatible solutions of the linear equations
(\ref{psieq}) such that  ${\bf g}=\Psi(\lambda =0)$ satisfies
the constraints eqs.(\ref{gcond}) and (\ref{gcond1}).

\section{The Dressing Method in Soliton Theory}
\subsection{The reduction group of the associated linear system}
\

We will consider now the symmetry group (or the so called
``reduction group'' \cite{zm1},
\cite{zm2}) enjoyed by  the linear system of equations
\begin{equation}
\Psi _{\xi}=\frac{U}{1-\lambda }\Psi ,\ \quad
\Psi _{\eta}=\frac{V}{1+\lambda }\Psi ,
\label{psieq1}
\end{equation}
when eqs.(\ref{gcond1}) hold.

It follows from the condition $\langle qJq\rangle =1$  that
the matrix ${\bf g}=I-2 q\rangle \langle gJ $ anticommutes with $U$
and $V$:
$$  {\bf g}U+U{\bf g}=0,\ \ \ {\bf g}V+V{\bf g}=0 . $$
This  implies that the matrix function ${\bf g}\Psi (1/\lambda ) $
satisfies the same equation as $\Psi(\lambda ) $ :
\begin{equation}
[g \Psi(1/\lambda )]_{\xi}=
\frac{U}{1-\lambda }[g \Psi(1/\lambda )] ,\ \ \
[g \Psi(1/\lambda )] _{\eta}=
\frac{V}{1+\lambda }[g \Psi(1/\lambda )] ,
\label{psieq2}
\end{equation}
Then, it can differ
from $\Psi(\lambda ) $ only
on a matrix multiplier which does not depend on $\xi, \eta $:
\begin{equation}
{\bf g}\Psi (1/\lambda ) = \Psi (\lambda )\delta_{1}(\lambda ) .
\label{gr1}\end{equation}

The vector $q\rangle$, the corresponding matrix
$\bf g$ and the currents $U,V$ are real.
Therefore $\Psi ^{*} ({\lambda ^*})$
is a solution of equations (\ref{psieq})
as well,  and we have
\begin{equation}
\Psi ^{*} ({\lambda ^*})=\Psi (\lambda )\delta _{2}(\lambda )
\label{gr2}\end{equation}
In addition, by using eq.(\ref{gr2}) twice, we find
\begin{equation}
\delta _{2}(\lambda )~\delta _{2}^*(\lambda^* ) = I
\label{grd2}\end{equation}

The fact that ${\bf g}\in SO(3,1)$ yields
$JU^T J=-U, JV^T J=-V$ and implies that $(J\Psi ^{\mbox{t}}(\lambda)
J)^{-1}$ obeys the same equation (\ref{psieq}) as $\Psi(\lambda ) $

\begin{equation}
(J\Psi ^{\mbox{t}}(\lambda) J)^{-1}=
\Psi(\lambda ) \delta _{3}(\lambda ).
\label{gr3}\end{equation}

The transformations (\ref{gr1}),(\ref{gr2}) and (\ref{gr3})
 generate a finite group
which is  called the reduction group of the problem and which
guarantees that  the properties (\ref{gcond1}) hold
for  ${\bf g}=\Psi(\lambda =0)$.

\subsection{Rational dressing}
\

Suppose we know a particular solution
${\bf g}_{(0)}(\eta ,\xi)$ of the string
equations (\ref{seq}).
We shall denote by $U_{(0)}(\eta ,\xi), V_{(0)}(\eta
,\xi)$ its corresponding currents (\ref{UV}),
and by $\Psi^{(0)} (\lambda, \eta ,\xi)$ , the corresponding
compatible solution of
the overdetermined system (\ref{psieq}).
We assume that $\Psi^{(0)}$
 as well as $U_{(0)}$ and $V_{(0)}$ are explicitly known.

To construct a new solution ${\bf g}$
we assume that the corresponding $\Psi $
function differs from $\Psi ^{(0)}$
on a rational matrix multiplier $\Phi (\lambda, \eta ,\xi )$
\begin{equation}
\Psi (\lambda )=\Phi (\lambda )\Psi^{(0)}(\lambda ).
\label{phipsi}\end{equation}
We assume that $\Phi (\lambda ) $  is  rational in $\lambda $, but, of
course it might have a complex dependence in  $\xi ,\eta$.
The dressing method consist on finding a matrix $\Phi(\lambda )$ such
that $\Psi (\lambda )$ given by eqs.(\ref{phipsi}) fulfils the linear
system (\ref{psieq1}) and the symmetry conditions (\ref{gr1})-(\ref{gr3}).
Then, once  $\Phi (\lambda ) $ is known, the string solution
${\bf g}(\eta ,\xi)$ follows from eq.(\ref{l0g}).

It follows from (\ref{l0g}),(\ref{gr1}) and (\ref{phipsi}) that
$\Phi $ should obey the following symmetries:
\begin{eqnarray}
& &\Phi (0)\Psi^{(0)}(0)\Phi (1/\lambda )=\Phi (\lambda ) \Psi^{(0)}(0),
\label{Gru1}\\
& & \Phi ^*(\lambda ^*)=\Phi (\lambda ),\label{Gru2}\\
& & J\Phi ^{\mbox{t}}(\lambda )J=\Phi ^{-1}(\lambda)
\label{Gru3}\end{eqnarray}

We assume that the constant matrices $\delta_{1}(\lambda
),\delta_{2}(\lambda )$ and $\delta_{3}(\lambda )$ coincide for the
dressed and the undressed solutions.

Suppose that the rational function $\Phi (\lambda )$ has a pole  at
the point $\lambda_{0}$. It follows from eqs.(\ref{Gru1})
and (\ref{Gru2}) that it
must have poles  at the points $1/\lambda _{0}, \lambda _{0}^* ,
 1/\lambda _{0} ^* $ as well and, in addition,  $\Phi (\infty )=I $.
Thus, the simplest (generic) possible case is
\begin{equation}
\Phi (\lambda )= I+\frac{A}{\lambda -\lambda _{0}}+
                \frac{A^*}{\lambda -\lambda _{0}^*}+
                \frac{B}{\lambda -\lambda _{0}^{-1}}+
                \frac{B^*}{\lambda -\lambda _{0}^{*-1}}
\label{phi}\end{equation}
where $A,B$ are matrix functions of $(\xi ,\eta )$
to be determined below.
This simplest case will be called
the {\it one-soliton solution} from now on.
We choose this name since in the context of non-linear integrable
equations in an infinite space interval (the sine-Gordon equation, for
instance), this minimal pole-structure  (the minimal number of poles
compatible with the symmetry group)
in $\lambda$ generates the {\it one soliton} solution (see for
example \cite{sol}).

Here we have taken into account
eqs.(\ref{Gru2}) and  (\ref{Gru1}). It follows from eq.
(\ref{Gru3}) that
\begin{equation}
\Phi ^{-1}(\lambda )= I+\frac{JA^{\mbox{t}}J}{\lambda -\lambda _{0}}+
                \frac{JA^\dagger J}{\lambda -\lambda _{0}^*}+
                \frac{JB^{\mbox{t}}J}{\lambda -\lambda _{0}^{-1}}+
                \frac{JB^\dagger J}{\lambda -\lambda _{0}^{*-1}}
\label{phiinv}\end{equation}
here \dag\  denotes Hermitian conjugation of a matrix.
The condition $\Omega (\lambda )=\Phi (\lambda ) \Phi ^{-1}(\lambda )=I$
can be imposed
in the following way: the right hand side
($\Omega (\lambda )$) is a
rational function of $\lambda $
which takes the value $I$ at the point $\lambda =\infty $,
then $\Omega (\lambda )$ will
be identically $I$ if it  does not have any
singularity on the Riemann sphere of $\lambda $. Double poles
would vanish if and only if
\begin{equation}
AJA^{\mbox{t}}=0,\ \ \ \ BJB^{\mbox{t}}=0.
\label{AA}\end{equation}
Thus the matrices $A,B$ are degenerated and
we can write them as a sum of bivectors
\begin{equation}
A= \sum_i a_i\rangle \langle x_iJ,
\ \ \ \ B=\sum_i b_i\rangle\langle y_iJ.
\label{ABI}
\end{equation}
The constraints  (\ref{AA}) imply
\begin{equation}
\langle x_i | J | x_j \rangle = 0 ~ {\rm for~ all~ pairs~ i,j}.
\label{pre}
\end{equation}
which  means that the vectors $x_i\rangle$ are null and mutually
pseudo-orthogonal. Therefore, since pseudo-orthogonal null vectors are
proportional, we have
$$
x_i\rangle = c_i ~ x\rangle
$$
and without loss of generality, we take:
\begin{equation}
A=a\rangle \langle xJ,\ \ \ \ B=b\rangle\langle yJ.
\label{AB}
\end{equation}
Now the constraints (\ref{AA}) read
\begin{equation}
\xl J\xr =0,\ \ \ \ \yl J\yr =0
\label{xjx}
\end{equation}

In addition, by requiring the residues of $\Omega (\lambda )$
to vanish at the points
$\lambda _{0},\, 1/\lambda _{0},\, \lambda _{0}^* ,\,
 1/\lambda _{0} ^*\, $ , we get

\begin{eqnarray*}
& &AJ +  JA^{\mbox{t}} +
AJ(\frac{B^{\mbox{t}}}{\lambda _0-\lambda _0^{-1}}+
\frac{A^{\dagger} }{\lambda _0-\lambda_0^*}+
\frac{B^{\dagger} }{\lambda _0-\lambda _0^{*-1}})+
(\frac{BJ}{\lambda _0-\lambda _0^{-1}}+
\frac{A^* J}{\lambda _0-\lambda_0^*}+
\frac{B^* J}{\lambda _0-\lambda _0^{*-1}})A^{\mbox{t}} =0\\
& &BJ  + JB^{\mbox{t}} +
BJ(-\frac{A^{\mbox{t}} }{\lambda _0-\lambda _0^{-1}}+
\frac{B^{\dagger} }{\lambda _0^{-1}-\lambda_0^{*-1}}+ \\
& &\frac{A^{\dagger} }{\lambda _0-\lambda _0^{*-1}})+
(-\frac{AJ}{\lambda _0-\lambda _0^{-1}}+
\frac{B^* J}{\lambda _0^{-1}-\lambda_0^{*-1}}+
\frac{A^* J}{\lambda _0-\lambda _0^{*-1}})B^{\mbox{t}}=0\\
& &A^* J  +  JA^{\dagger} + A^*
J(\frac{B^{\dagger} }{\lambda _0^*-\lambda _0^{*-1}}+
\frac{A^{\mbox{t}} }{\lambda ^* _0-\lambda_0}+
\frac{B^{\mbox{t}} }{\lambda _0^*-\lambda _0^{-1}})+
(\frac{B^* J}{\lambda _0^*-\lambda _0^{*-1}}+
\frac{A J}{\lambda _0^*-\lambda_0}+
\frac{B J}{\lambda _0^*-\lambda _0^{-1}})A^{\dagger}=0\\
& & B^* J +  JB^{\dagger} + B^*
J(-\frac{A^{\dagger} }{\lambda^*_0-\lambda _0^{*-1}}+
\frac{B^{\mbox{t}} }{\lambda _0^{*-1}-\lambda_0^{-1}}+
\frac{A^{\mbox{t}}}{\lambda _0^*-\lambda _0^{-1}})+ \\
& &(-\frac{A^* J}{\lambda _0^*-\lambda _0^{*-1}}+
\frac{B J}{\lambda _0^{*-1}-\lambda_0^{-1}}+
\frac{A J}{\lambda _0^*-\lambda _0^{-1}})B^{\dagger} =0
\end{eqnarray*}

Later on we shall demonstrate that the periodicity condition
on $\sigma $ can be satisfied  only in the
case where all  poles ($\lambda _0,
\lambda _0^*, \lambda_0^{-1}, \lambda _0^{*-1} $)
 of $\Phi (\lambda )$ are  purely  imaginary
[see eqs.(\ref{lammu}),(\ref{alfa}) and (\ref{pita})].
 From now on, we shall denote $\lambda _0=i \kappa , \kappa \in R$.
Substituting the bivectoral
representation (\ref{AB}) in the above equations and by separating
bivectors we get the system of vector equations with respect
to $a\rangle, b\rangle$
\begin{eqnarray*}
\br \frac{i \xl J\yr}{\kappa+\kappa ^{-1}}+
\car \frac{i \cxl J\xr}{2\kappa }+
\cbr \frac{i \cyl J\xr}{\kappa-\kappa^{-1}}&=&\xr\\
\cbr \frac{i \cxl J\cyr}{\kappa+\kappa ^{-1}}+
\ar \frac{i \cxl J\xr}{2\kappa }+
\br \frac{i \cxl J\yr}{\kappa-\kappa^{-1}}&=&-\cxr\\
-\ar \frac{i \xl J\yr}{\kappa+\kappa ^{-1}}-
\cbr \frac{i \cyl J\yr}{2\kappa ^{-1}}+
\car \frac{i \cxl J\yr}{\kappa-\kappa^{-1}}&=&\yr\\
-\car \frac{i \cxl J\cyr}{\kappa+\kappa ^{-1}}-
\br \frac{i \cyl J\yr}{2\kappa ^{-1}}+
\ar \frac{i \cyl J\xr}{\kappa-\kappa^{-1}}&=&-\cyr
\end{eqnarray*}
By bivectoral separations we mean the following trick:
suppose ve have an equation of the form
$ p\rangle \langle X+X\rangle\langle p=0$ where $p\rangle ,X\rangle$
are vectors and $p\rangle\not = 0$. Then the only solution of this
equation is $X\rangle =0$ and the matrix equation is reduced to a
vector one.

One can solve this system of linear equations and express the
vectors $\ar, \br$ in terms of  $\xr ,\yr $
\begin{eqnarray}
\ar &=& \frac{2i\kappa (\kappa ^4 -1)}
{\delta}[(\kappa ^4 -1)\cxr \cyl J\yr
+2 (1+\kappa ^2)\cyr \cxl J\yr +
2 (1-\kappa ^2) \yr \cxl J\cyr ]\label{ar}\\
\br &=& \frac{2i (\kappa ^4 -1)}{\kappa \delta }
[(\kappa ^4 -1)\cyr \cxl J\xr
+2 \kappa ^2(1+\kappa ^2)\cxr \cyl J\xr +
2 \kappa ^2(1-\kappa ^2) \xr \cxl J\cyr ]
\label{br}\end{eqnarray}
where $\delta$ is the scalar function
\begin{equation}
\delta =(1-\kappa ^4)^2 \cxl J \xr \cyl J \yr +
4 \kappa ^2 (1+\kappa ^2 )^2 \cxl J\yr\cyl J\xr-
4\kappa ^2 (1-\kappa ^2)^2 \cxl J \cyr \xl J\yr
\label{delta}
\end{equation}
At the moment we have  fulfiled the reduction constraints
(\ref{Gru2}), (\ref{Gru3})
completely, but the constraint eq.(\ref{Gru1}) has not yet been
imposed. One can prove without loose of generality,
that eq.(\ref{Gru1}) is verified,
 when the vectors $\xr,\yr$, fulfil
\begin{equation}
\yr = \Psi ^{(0)}(0)\; \xr .
\label{xy}\end{equation}

Assembling  eqs.(\ref{AB}) and eqs.(\ref{ar})-(\ref{xy})
all together one can find $\Phi$ (eq.(\ref{phi})) as a function
of $x\rangle , \lambda , \lambda _{0}$
\begin{equation}
\Phi =\Phi (\lambda ,\lambda _{0},x\rangle ) \label{phi0}
\end{equation}
Now, we are interested
in the value of this function at $\lambda =0$,
since it gives a new solution
$\g =\Phi (0)\g_{(0)}  $. One can check that
\begin{eqnarray}
\g & =& \g_{(0)} -\frac{4 (1-\kappa ^4)}{\delta}[(1-\kappa ^4)
(F\g_{(0)}  +\g_{(0)}  F)\cxl J\xr +\label{otvet}\\
& & 2 (1+\kappa ^2)(\kappa ^2 F -\g_{(0)}
F\g_{(0)} )\cxl J\g_{(0)} \xr-
2 (1-\kappa ^2) \mbox{Re}
((\kappa ^2 H +\g_{(0)}  H\g_{(0)} )\cxl J\g_{(0)} \cxr)],
\nonumber
\end{eqnarray}
where
$$
F=\mbox{Re}(\cxr \xl J),\ \ \ \ H=\xr \xl J
$$

Thus, we have parametrised the new solution $\g $
by a real number $\kappa $ and a complex vector $\xr $
of zero pseudolength. Since $\g $ satisfies the conditions
(\ref{gcond1}), it has the form of eq.(\ref{G}) and the corresponding
vector $q\rangle$, can be found by projecting eq.(\ref{G}) on an
arbitrary constant vector $ p\rangle$. (For instance $(1,0,0,0)$).
We find in this way:
\begin{equation}
q\rangle = \frac{p\rangle -\g p\rangle}{\sqrt{2(\langle pJp\rangle -
\langle pJ\g p\rangle )}},
\label{newq}\end{equation}

Now, to construct a solution $q\rangle$ of the string equations
(\ref{seq}, \ref{scond}), one needs only
to determine the dependence of
$\xr $ and $\kappa $ on the variables $\xi , \eta $.

\subsection{Evolution of  $\xr$ and $\kappa $ in
$\xi $ and $\eta $.}
\

Now the problem is to find  the evolution of $\xr $
 and $\kappa $ in $\xi $ and $\eta $.  It follows from eqs.(\ref{psieq}),
(\ref{phipsi}) that
\begin{eqnarray}
\Phi _{\xi}+\frac{1}{1-\lambda }\Phi U_{(0)}&
=&\frac{1}{1-\lambda }U\Phi ,\label{phixi}\\
\Phi _{\eta}+\frac{1}{1+\lambda }\Phi V_{(0)}&
=&\frac{1}{1+\lambda }V\Phi ,
\label{phieta}\end{eqnarray}
where $U,V$ are still undetermined functions of $\eta , \xi $ which do
not depend on $\lambda $. Let us rewrite equations (\ref{phixi}),
(\ref{phieta}) in the form
\begin{eqnarray}
\Phi (-\partial_{\xi}+\frac{U_{(0)}}{1-\lambda })\Phi ^{-1}&
=&\frac{U}{1-\lambda } ,\label{phixi1}\\
\Phi (-\partial_{\eta}+\frac{V_{(0)}}{1+\lambda })\Phi ^{-1}&
=&\frac{V}{1+\lambda }.
\label{phieta1}\end{eqnarray}
Consider the l.h.s. of (\ref{phixi1}). It is a rational function of
$\lambda $ with a pole at $\lambda = 1$
and at $\lambda \in \{\lambda _{0},\,
\lambda _{0}^{-1}\, , \lambda _{0}^{*}\, , \lambda _{0}^{*-1}\} $ ,
 but the r.h.s.
has only one pole at  $\lambda = 1$ .
Thus to fit  eq.(\ref{phixi1}) we have to
set the residues at $\lambda _{0}, \, \lambda _{0}^{-1}, \,
\lambda _{0}^{*}\, $ and $\lambda _{0}^{*-1}\, $ equal to zero.
In fact, it is
sufficient to require the vanishing of  the residue
 at $\, \lambda _{0}\,$ only. All other
 residues will vanish due to the action of the reduction group.
The condition
\begin{equation}
\mbox{res}|_{\lambda _{0}}\Phi
(-\partial_{\xi}+\frac{U_{(0)}}{1-\lambda })\Phi ^{-1}=0
\label{xires}\end{equation}
yields
\begin{equation}
A (\partial_{\xi}-
\frac{U_{(0)}}{1-\lambda _{0}})JA^{\mbox{t}}=0 \label{A1}
\end{equation}
and
\begin{eqnarray}
& & A  (\partial_{\xi}-\frac{U_{(0)}}{1-
\lambda _{0}})J(\frac{B^{\mbox{t}}}{\lambda _0-\lambda _0^{-1}}+
\frac{A^{\dagger} }{\lambda _0-\lambda_0^*}+
\frac{B^{\dagger} }{\lambda _0-\lambda _0^{*-1}})+\nonumber\\
& & (\frac{B}{\lambda _0-\lambda _0^{-1}}+
\frac{A^* }{\lambda _0-\lambda_0^*}+
\frac{B^* }{\lambda _0-\lambda _0^{*-1}})
(\partial_{\xi}-\frac{U_{(0)}}{1-
\lambda _{0}}) JA^{\mbox{t}} =0 \label{A2}
\end{eqnarray}
Both equations will be satisfied if
\begin{equation}
 (\partial_{\xi}-\frac{U_{(0)}}{1-\lambda _{0}})\, x\rangle =0
\label{xeq}
\end{equation}
Thus, the simultaneous solution of eqs.(\ref{A1}) and (\ref{A2}) is
\begin{equation}
x\rangle =\Psi^{(0)}(\eta ,\xi ;\lambda _0 )\, x_{0}\rangle
\label{xxi}
\end{equation}
where $x_{0}\rangle $ is any complex constant vector of zero
pseudolength,  and $\lambda _{0}$ turns not to depend on $\eta , \xi $.

Moreover, the  solution  (\ref{xxi}) of equation (\ref{xires}) is also a
solution of the equation
\begin{equation}
\mbox{res}|_{\lambda _{0}}\Phi
(-\partial_{\eta}+\frac{V_{(0)}}{1+\lambda })\Phi ^{-1}=0 .
\label{etares}\end{equation}

Finally, the $\eta ,\xi $ dependance of the vector $x\rangle $ is given
by eq.(\ref{xxi}).
Together with eq.(\ref{otvet}) it gives the one soliton solution.

The wave function $\Psi(\eta ,\xi ;\lambda)$ corresponding to the
one soliton solution (let us denote it by $\Psi_{1} (\eta ,\xi ;\lambda)$
can be regarded as a function of $\lambda ,\lambda _{0},
\Psi ^{(0)}(\eta ,\xi ,\lambda )$, and $x_{0}\rangle $ (see
eqs.(\ref{phipsi}) and (\ref{phi0})).
\begin{equation}
\Psi _{1}(\eta ,\xi ;\lambda)=\Phi(\lambda , \lambda _{0},
J\Psi ^{(0)}(\eta ,\xi ;\lambda _{0})Jx_{0}\rangle )
\Psi _{0}(\eta ,\xi ;\lambda )
\label{1spsi}
\end{equation}
The wave function corresponding to a $n$-soliton solution
can be obtained recursively through the relation
\begin{equation}
\Psi _{n}(\eta ,\xi ;\lambda)=\Phi(\lambda , \lambda _{n-1},
J\Psi _{n-1}(\eta ,\xi ;\lambda _{n-1})Jx_{n-1}\rangle )
\Psi _{n-1}(\eta ,\xi ;\lambda ) \, ,
\label{nspsi}
\end{equation}
and the corresponding solution of the chiral model
$\g =\Psi _{n}(\eta ,\xi ;\lambda = 0)$
will satisfy all the reduction conditions eqs.(\ref{gr1}),(\ref{gr2})
 and (\ref{gr3}).

\section{The choice of the background solution}
\

Let us construct now explicit solutions by applying the above
procedure . To begin with we shall consider a three dimensional de Sitter
spacetime ($D=3$).

As a background  starting solution we choose
for simplicity the solution
$q_{(0)}(\sigma,\tau)$ found in ref.\cite{dms}. This solution
corresponds to the trivial $\alpha = 0$ solution of the sinh-Gordon
equation and it  is given by,
\begin{equation}
q_{(0)}= \frac{1}{\sqrt{2}}\left( \begin{array}{r}
\sh \\ \ch \\ \co
\\
\si\end{array}\right), \ \quad
q_{(0)\xi }= \frac{1}{\sqrt{2}}\left( \begin{array}{r}
\ch \\ \sh\\  \si\\
- \co \end{array}\right),
\label{qu0}
\end{equation}

\begin{equation}
q_{(0)\eta }= \frac{1}{\sqrt{2}}\left( \begin{array}{r}
\ch \\ \sh\\ -\si \\ \co \end{array}\right), \ \quad
b_{(0)}= \frac{1}{\sqrt{2}}\left( \begin{array}{r}
\sh \\ \ch \\ - \co\\-\si \end{array}\right),
\label{quu}
\end{equation}

For this solution, we have
$$
q_{(0)\xi \xi }=q_{(0)\eta \eta }=b_{(0)},
\ \ \ \ q_{(0)\xi \eta }=q_{(0)},\ \ \ \
b_{(0)\xi }=q_{(0)\eta },\ \ \ \ b_{(0)\eta }=q_{(0)\xi }.
$$

$$
\langle q_{(0)\xi }Jq_{(0)\eta }\rangle =-1,\ \ \ \
\langle q_{(0)}Jq_{(0)}\rangle =1,\ \ \ \
\langle b_{(0)}Jb_{(0)}\rangle =1, \ \ \ \
\mbox{other   }\langle\cdot J \cdot \rangle=0
$$

$$
U_{(0)}=2 q_{(0)\xi }\rangle\langle q_{(0)} J -2 q_{(0)}\rangle
\langle q_{(0)\xi }J
, \ \ \ \ \
V_{(0)}=2 q_{(0)\eta }\rangle\langle q_{(0)} J-2 q_{(0)}\rangle
\langle q_{(0)\eta }J
$$
where $J$ is given by eq.(\ref{l1}).
Let us define
\begin{equation}
Q(\xi, \eta)=(q_{(0)\xi }, q_{(0)\eta }, q_{(0)}, b_{(0)}) \ \ ,
\label{QQQ}
\end{equation}
then we find by direct calculation that
$$
Q^T JQ=\left(\begin{array}{rrrr}
0,& - 1,& 0, & 0 \\
-1, & 0, & 0, & 0 \\
0, & 0, & 1, & 0 \\
0, & 0, & 0, & 1 \end{array}\right)
$$

$$
U_{(0)}Q=(0,2q_{(0)},2q_{(0)\xi },0),\ \ \ \
V_{(0)}Q=(2q_{(0)},0,2q_{(0)\eta },0)
$$
and

$$
Q^T JU_{(0)}Q=\left(\begin{array}{rrrr}
0,& 0,& 0, & 0 \\
0, & 0, & -2, & 0 \\
0, & 2, & 0, & 0 \\
0, & 0, & 0, & 0 \end{array}\right)
, \quad
Q^T JV_{(0)}Q=\left(\begin{array}{rrrr}
0,& 0,& -2, & 0 \\
0, & 0, & 0, & 0 \\
2, & 0, & 0, & 0 \\
0, & 0, & 0, & 0 \end{array}\right)
$$

$$
Q^T JQ_{\xi }=\left(\begin{array}{rrrr}
0,& 0,& 0, & -1 \\
0, & 0, & -1, & 0 \\
0, & 1, & 0, & 0 \\
1, & 0, & 0, & 0 \end{array}\right)
,\quad
Q^T JQ_{\eta}=\left(\begin{array}{rrrr}
0,& 0,& -1, & 0 \\
0, & 0, & 0, & -1 \\
1, & 0, & 0, & 0 \\
0, & 1, & 0, & 0 \end{array}\right)
$$

We have to solve two compatible equations for  $\Psi $
\begin{equation}
\Psi ^{(0)}_{\xi}=\frac{U_{(0)}}{1-\lambda }\Psi ^{(0)} ,\ \quad
\Psi ^{(0)}_{\eta}=\frac{V_{(0)}}{1+\lambda }\Psi ^{(0)}.
\label{psi0eq}
\end{equation}
Let us make the gauge transformation
\begin{equation}
\Psi ^{(0)} = Q(\xi, \eta) ~~ \Xi(\xi, \eta) ,
\label{deft}
\end{equation}
then
\begin{equation}
Q^T J Q_{\xi} \Xi + Q^T J \Xi_{\xi} =
\frac{Q^T J U_{(0)} Q}{1-\lambda} \Xi
\ \ , \ \
Q^T J Q_{\eta} \Xi +
Q^T J Q \Xi_{\eta} = \frac{Q^T J V_{(0)} Q}{1+\lambda} \Xi \ ,
\label{trga}
\end{equation}
and $\Xi(\xi, \eta) $ satisfies the following
equations with {\it constant} coefficients

\begin{eqnarray}
\Xi_{\xi } = \left(\begin{array}{rrrr}
0,& 0,& \mu^{-2}, & 0 \\
0, & 0, & 0, & -1 \\
0, & \mu^{-2}, & 0, & 0 \\
-1, & 0, & 0, & 0 \end{array}\right) \Xi(\xi, \eta)
\label{treq}\end{eqnarray}
\begin{eqnarray}
\Xi_{\eta } = \left(\begin{array}{rrrr}
0,& 0,& 0, & -1 \\
0, & 0, & \mu^2, & 0 \\
\mu^2, & 0, & 0, & 0 \\
0, & -1, & 0, & 0 \end{array}\right) \Xi(\xi, \eta) \\
\label{truq}\end{eqnarray}

The solution will be defined on the Riemann surface $\Gamma$
which covers twice the complex plane $\lambda$:
\begin{equation}
\mu^2 = \frac{1-\lambda}{1+\lambda}
\label{lammu}
\end{equation}
The points $ \lambda = \pm 1$ are the branching points of $\Gamma$.

The fundamental set of solutions of eqs.(\ref{treq})-(\ref{truq}) is
given by:

\begin{eqnarray}
\exp(\mu\eta+\mu^{-1} \xi)\left(\begin{array}{r}
-\mu^{-1} \\
-\mu \\
-1 \\
1 \end{array}\right) , ~~
\exp(-\mu\eta-\mu^{-1} \xi)\left(\begin{array}{r}
\mu^{-1} \\
\mu \\
-1 \\
1 \end{array}\right), \\
\exp(i\mu\eta-i \mu^{-1}\xi)\left(\begin{array}{r}
i\mu^{-1} \\
-i\mu \\
1 \\
1 \end{array}\right), ~~
\exp(-i\mu\eta+i\mu^{-1} \xi)
\left(\begin{array}{r}
-i\mu^{-1} \\
i\mu \\
1 \\
1 \end{array}\right).
\label{solf}\end{eqnarray}

It will be convenient to use the following linear combinations
of the above solutions:
\begin{equation}
\Xi(\mu, \xi, \eta ) =  \Lambda(\mu) . \Pi(\mu, \xi, \eta)
\label{lapi}
\end{equation}
where:
\begin{equation}
\Lambda(\mu) = {\rm diag}(\mu^{-1}, \mu, 1, 1 )
\label{lamu}
\end{equation}

\begin{equation}
\Pi(\mu, \xi, \eta) = \frac{1}{\sqrt{2}}  \left(\begin{array}{rrrr}
\cosh\gamma ,& -\sinh\gamma,& \Si, & -\Co \\
\cosh\gamma , & -\sinh\gamma, & -\Si, & \Co \\
\sinh\gamma, & -\cosh\gamma , & -\Co, & -\Si \\
-\sinh\gamma, & \cosh\gamma , & -\Co, & -\Si \end{array}\right)
\label{mafi}
\end{equation}

and

\begin{equation}
\theta = \mu\eta-\frac{1}{\mu} \xi ~ , ~ \gamma =
\mu\eta+\frac{1}{\mu} \xi \\
\label{tega}
\end{equation}

Thus, $\Psi^{(0)}(\mu, \xi, \eta)$ as expressed by
\begin{equation}
\Psi^{(0)}(\mu, \xi, \eta) = Q(\xi, \eta) .
\Lambda(\mu) . \Pi(\mu, \xi, \eta)
\label{qlpi}
\end{equation}
is the solution of the system (\ref{psi0eq}) fulfilling the constraint
(\ref{l0g}) for the solution $q_{(0)}$. That is :
$$
\Psi^{(0)}(\mu = 1, \xi, \eta) = 1 - 2 |q_{(0)}><q_{(0)}|J .
$$

We want solutions periodic in $\sigma$ with period $2\pi$.
We see from eq.(\ref{lapi})-(\ref{qlpi}) that we have hyperbolic
functions on the argument
\begin{equation}
\frac{\sigma}{2} (\mu - \mu^{-1}) + \frac{\tau}{2}(\mu + \mu^{-1})
\label{argu}
\end{equation}
and trigonometric functions with the argument
$$
\frac{\sigma}{2} (\mu + \mu^{-1}) + \frac{\tau}{2}(\mu - \mu^{-1})
$$
The solution to the $\sigma$-periodicity condition requires to have
\begin{equation}
\mu = \exp[i\alpha]
\label{alfa}
\end{equation}
with real $\alpha$ and  $\cos\alpha$ and $\sin\alpha$ to be
rational numbers. The general solution is given by the
Pythagorean numbers:
\begin{equation}
\cos\alpha = \frac{m^2-n^2}{m^2+n^2} ~,~ \ \
\sin\alpha = -\frac{2mn}{m^2+n^2} ~,~ \ \  m,n~=~{\rm ~ integers}
\label{pitg}
\end{equation}
That is,
\begin{equation}
\mu = \frac{m+in}{m-in}~,~~ m,n~=~{\rm ~ integers}
\label{pita}
\end{equation}
We get in this way solutions with period $2\pi (m^2 + n^2 ) $ in
$\sigma$. Upon the rescaling:
\begin{equation}
\sigma \to \sigma (m^2 + n^2) ~ ,~~ \tau \to \tau (m^2 +
n^2) ,
\label{escal}
\end{equation}
we set the period to $2\pi$. Notice that now the background solution
$q_{(0)}(\sigma,\tau)$ will be winded $(m^2 + n^2)$ times
around the origin in de Sitter space.

\section{The soliton string solutions and their properties}
\

We have now all the elements to obtain the explicit expression for
the solution $|q(\eta,\xi)\rangle$ of the string
eqs.(\ref{seq})-(\ref{scond}). The explicit expression for the solution $
\Psi ^{(0)}(\eta ,\xi ;\mu_0 )$ given by eq.(\ref{qlpi}) can be directly
obtained by computing the indicated matrix multiplication; $ Q(\eta
,\xi)$  is given by eqs.(\ref{qu0}-\ref{QQQ}); $\Lambda(\mu)$ and
$\Pi(\mu, \xi, \eta)$ are given by eqs.(\ref{mafi}). This was done
with the help of the computer program of symbolic
calculation ``Mathematica''.

By projecting $\Psi^{(0)}(\eta ,\xi ;\mu_0 )$ thus obtained
on a constant and complex null vextor $|x_0\rangle$ , we
have directly the vector $|x\rangle$. The matrix $g(\xi, \eta)$ is
obtained from eq.(\ref{otvet}) also using ``Mathematica''.
Finally, the explicit solution  $|q(\eta,\xi)\rangle$ is obtained
by inserting $g(\xi, \eta)$ in eq.(\ref{newq}). These string solutions
depend on one complex parameter $\mu$ that depends on two integers
$n$ and $m$ (see eq.(\ref{pita})), and one complex null vector
$x_0\rangle$ , that is, three complex independent numbers. Only two
independent complex components
remain in fact since $g(\xi, \eta)$ is homogeneous in $x_0\rangle$.
As can be seen in eq.(\ref{otvet}) ,
the change $x_0\rangle \to c~ x_0\rangle$
 where $c$ is a complex number, leaves the solution invariant.
The dependence on  $x_0\rangle$ is precisely like what happens
for strings in D-dimensional Minkowski space-time, in which the solution
depends on $2(D-2)$ complex coefficients. They account for the
$(D-2)$-transverse degrees of freedom and for the two helicity modes
(right and left movers) . Here, we are in three space-time
dimensions and so we obtain two complex coefficients
corresponding to the transverse degrees of freedom.

It can be noticed, that the linear system (\ref{psieq1}) fulfilled by
$\Psi(\eta ,\xi, \lambda)$ is invariant under conformal
transformations on $\xi, \eta$ .
Thus the dressing transformations do not generate conformal modes but
only physical (transverse) modes.

The vector  $x_0\rangle$ describes
the polarization of the string; the integers $m, n$ associated to
the $\sigma$ periodicity, label the string modes. In Minkowski
space-time,  only one integer labels the right-modes  and another one
labels the  left-modes. Here, we obtain two independent integers
for each mode. Notice that our modes combine left and right movers in
a non-linear and precise way.

The resulting solution $q(\sigma,\tau) = (q^0,q^1,q^2,q^3) $ is
a complicated combination of trigonometric functions of $\sigma$
and hyperbolic functions of $\tau$ . From
eqs.(\ref{argu})-(\ref{pita}), we see that we have trigonometric
functions on the arguments
$$
\sigma \; \frac{2mn}{m^2+n^2} \quad {\rm and} \quad \sigma \;
\frac{m^2-n^2}{m^2+n^2}
$$
and hyperbolic functions on the arguments
\begin{equation}
\tau \; \frac{2mn}{m^2+n^2} \quad {\rm and} \quad \tau \;
\frac{m^2-n^2}{m^2+n^2}
\label{freta}
\end{equation}
That is, these string-solitonic solutions,
{\bf do not oscillate in time}. This is a typical feature
of string unstability \cite{agn},\cite{gsv} which is present for
strings in inflationary type backgrounds, i. e.
 accelerated expanding like de Sitter, and in black holes.
The new feature here is that the string does not oscillate
in time, neither for $\tau \to 0$ nor for $\tau \to \pm \infty $.
It can be noticed that in decelerated expanding backgrounds,
as it is the case in the standard Friedman-Robertson-Walker (FRW)
expansion, string instability does not occur and the string behaviour
is oscillating in $\tau$ \cite{gsv}. This was recently confirmed for
all values of $\tau$, in the FRW universe, where explicit string
solutions has been found \cite{ani}.
The non-oscillatory behavior in time can be understood from the fact
that the string motion in de Sitter spacetime reduces to a sinh-Gordon
equation with negative potential \cite{prd}. In $D = 3$, this is
precisely the standard sinh-Gordon equation, whose potential
unbounded from below (see fig. 1) is responsible of the instability.
By defining
\begin{equation}
e^{\alpha(\sigma,\tau)} = - q_{\xi}.q_{\eta} ~,
\label{tama}
\end{equation}
the string equations (\ref{seq}) and string constraints (\ref{scond})
in de Sitter spacetime can be reduced to the sinh-Gordon equation
\begin{equation}
\alpha_{\tau \tau} -\alpha_{\sigma\sigma} -
e^{\alpha} + e^{-\alpha} = 0
\label{senh}
\end{equation}
Therefore, in order to find a solution in $D=3$ de Sitter spacetime,
one can start from a $\sigma$-periodic solution of the sinh-Gordon
equation (\ref{senh}) and insert it in the string equations
(\ref{seq}):
\begin{equation}
\left[\partial^2_{\tau} -\partial^2_{\sigma} - e^{\alpha(\sigma,\tau)}
\right]q(\sigma , \tau) = 0
\label{klgo}
\end{equation}
Then, one must solve the {\bf linear} equation (\ref{klgo})
in $q(\sigma , \tau)$ and impose the constraints (\ref{l1})
and (\ref{scond}). This is actually an alternative method
to the dressing method, to obtain string
solutions in de Sitter spacetime.

The function $e^{\alpha(\sigma,\tau)}$ has a clear physical
interpretation, as it determines the proper string size. The invariant
interval between two points on the string, computed with the spacetime
metric, is given by
\begin{equation}
ds^2 = \frac{1}{H^2} dq.dq\; =
\; \frac{1}{2 H^2}\; e^{\alpha(\sigma,\tau)}\;
( d\sigma^2 -d \tau^2)
\label{inin}
\end{equation}
The energy density of the sinh-Gordon model here
\begin{equation}
{\cal H} = \frac{1}{2}~[~(\frac{\partial \alpha}{\partial \tau})^2+
(\frac{\partial
\alpha}{\partial \sigma })^2~] - 2\cosh\alpha(\sigma,\tau)~,
\label{dene}
\end{equation}
determines the potential
\begin{equation}
V_{eff} = - 2 \cosh \alpha   \quad  ,
\label{pote}
\end{equation}
This potential has absolute minima at $\alpha = +\infty$
and  $\alpha = -\infty$. As the time $\tau$ evolves,
$\alpha(\sigma,\tau)$ generically approach one of these infinite
minima. The first minimun corresponds to an infinitely large string
whereas the second one describes a collapsed configuration. That is,
the string in de Sitter spacetime will tend generically either
to inflate (when $\alpha \to +\infty$) or to collapse to a point
(when $\alpha \to -\infty$).

The background string solution $q_{(0)}(\sigma,\tau)$ given by
eq.(\ref{qu0}) , corresponds to the sinh-Gordon solution $\alpha = 0$.
This means a string with {\bf finite constant} proper size (equal to
$1/H$ ). In the sinh-Gordon model this corresponds
to a particle at the maximun of the potential
$V_{eff} = - 2$ and with zero velocity.

Let us recall that for a given time $q_0$, the de Sitter space is
a sphere $S^2$ with radius $R = {1 \over H}\sqrt{1 + q_0^2}$ .
For the background solution $q_{(0)}(\sigma,\tau)$ given by
eq.(\ref{qu0}), we have $R(\tau) = {1 \over H}\sqrt{1 +
{1 \over 2}\sinh^2\tau} $.
As de Sitter universe expands for $\tau \to
\infty$ , the string size  $e^{\alpha(\sigma,\tau)} = 1$ remains
here constant. This solution is probably unstable under small
perturbations.

It must be noticed that the integers $(m,n)$ of the solitonic
solutions  (\ref{pitg}) have the meaning of string winding.
They label the different ways in which the string wind in
the spatial compact dimensions (here $S^2$). Notice that our
string solutions  do not oscillate in time in spite of the fact
that we are in a lorentzian signature spacetime. (The dependence
on $\tau$ is hyperbolic).

In figs. 3 and 4 we plot the one soliton solutions
$|q(\sigma,\tau)\rangle =
(q^0,q^1,q^2,q^3) $ found here. They show the three dimensional
spatial projections $ (q^1,q^2,q^3) $ as a function of $\sigma$,
for a given polarization vector $x^0\rangle $, different values of
$m, n$ and two differents values of $\tau$.
Figs. 3a and 3b show the same solution
($n = 2, m = 1,\;  x^0\rangle = (1, -1 , .1 , .1i ) $) for
two different values of $\tau$. Figs. 4a and 4b show the evolution
for a higher winding number ($n=5$), and polarization vector
$x^0\rangle = (1, -1 , 1 , i ) $.

For comparison, let us recall, that in $D = 2$ spacetime dimensions,
in which the string motion reduces to the Liouville equation, the
{\it exact general} solution is a string wound n times around the
de Sitter space and evolving with it. The string covers n times de
Sitter space which is here a circle $S^1$. This solution is
given by \cite{prd}
\begin{eqnarray}
q^0 = - \cot{n\tau} ~~ , \quad q^1 = \frac{\cos n\sigma}{\sin
n\tau} ~~,\quad  q^2 = \frac{\sin n\sigma}{\sin n\tau} \\ ~
\nonumber ~ \\
0 < \sigma \leq 2\pi~,~~~0 < \tau \leq \pi/n
\label{enrl}
\end{eqnarray}

The invariant interval between two points of the string
\begin{equation}
ds^2 = \frac{1}{H^2 \sin^2 n\tau} ( d\sigma^2 -d \tau^2)
\label{in2d}
\end{equation}
exhibits the typical feature of string instability : in the
asymptotic regions $\tau \to 0^+ $ and $\tau \to \pi/n $,
the proper string length blows up.
We also see that the string does not have ``enough time''
to oscillate in one expansion time of the universe :
the oscillation period of the string {\bf coincides}
with the expansion time of the
universe. When the string accomplishes one oscillation, the universe
has ended.

In order to analyze the exact solutions  $|q(\sigma,\tau)\rangle$ ,
it is convenient to use the coordinates $(T, X^1, X^2)$ in this
$2+1$-dimensional de Sitter spacetime:
\begin{eqnarray}
q^0 = ~\sinh{H T}~ + \frac{H^2}{2} \exp(H T)\, \left[ \,(X^1)^2  +
 (X^2)^2 \,\right] \\
q^1 = ~\cosh{H T}~ - \frac{H^2}{2} \exp(H T)\, \left[\, (X^1)^2  +
 (X^2)^2 \,\right] \\
q^{2} = ~H \exp(H T)~ X^{1} ~~,\qquad
q^{3} = ~H \exp(H T)~ X^{2} ~~,\\
 -\infty < T , ~ X^{1}, X^2 < + \infty . \nonumber
\label{traf}
\end{eqnarray}
That is,
\begin{equation}
T = \frac{1}{H}\log(q^0 + q^1) ~~,~~~ X^1 =\frac{1}{H}
\frac{q^2}{q^0 + q^1} ~~,~~X^2 =\frac{1}{H}
\frac{q^3}{q^0 + q^1} ~~
\label{otraf}
\end{equation}
The cosmic time $T$ and the conformal time $\eta$ are related by
\begin{equation}
\eta = - \frac{1}{H}~~ e^{-HT}~~,-\infty < \eta \leq 0~,
\label{etat}
\end{equation}
in terms of which, the line element takes the form
\begin{equation}
ds^2 = -dT^2 + e^{2HT}\left[ (dX^1)^2 + (dX^2)^2 \right] =
\frac{1}{H^2\eta^2}\left[ -(d\eta)^2 + (dX^1)^2 + (dX^2)^2 \right].
\label{dese}
\end{equation}
We now analyze the properties and new features
exhibited by these solutions.

First of all, let us analyze the cosmic time coordinate
$T = T(\sigma,\tau)$. We have studied $T$ as a function of
$\tau$ for different fixed values
of $\sigma$ and viceversa. These functions have been obtained
numerically for a wide family of solutions labeled by
different values of the parameters $n, m$ and $|x^0\rangle$ . We
report here only two significative cases, which show the {\it generic}
features, irrespective of the particular values of these parameters.

Fig. 5 shows $ T $ as a function of $\tau$ for the values of $\sigma$
indicated in the picture. We depict $ T $ for $n = 4 , m = 1$ and
a generic $|x^0\rangle = ( 1 + i, .6 + .4 i , .3 + .5 i ,
.77 + .79 i ) $.

In fig. 6 we depict  $T$ as a function of $\tau$ for the solution with
$n = 4 , m = 1 $ and  $|x^0\rangle = ( 1 , -1  , i  , 1 ) $.
We see that our solution in the generic case
describes actually {\bf five} strings,
as it can be seen from the fact that for a given value of $T$
we find five different values of $\tau$ .That is, $\tau$ is a
{\bf multivalued} function of $T$ for any fixed $\sigma$ . This is
an entirely new feature for strings in curved spacetime.
It has no analogy in flat spacetime where the time coordinate obeys
$ \left[\partial^2_{\tau} -\partial^2_{\sigma}\right]T = 0 $ ,
and therefore, using the conformal transformations
$$
\sigma \pm \tau = f_{\pm}(\sigma' \pm \tau')
$$
allows to choose the light-cone gauge, in which $T$ (or a
null-like combination of it) is proportional to $\tau$. In curved
spacetime, this is not possible in general.
Only in some geometries like shock-waves or gravitational
plane waves, the light-cone choice is possible for all
$\tau$ \cite{eri}.
In asymptotically flat spacetimes this choice is only possible
asymptotically \cite{agn}.
When $\tau$ is a univalued function of the time coordinate,
the solution of the
string equations describes {\it only one} string. This is
the case in geometries where the the light-cone choice of gauge
is possible for all $\tau$ .
In spacetimes, as de Sitter,  where $\tau$ is a multivalued
function of the time  coordinate, the solution of the string
equations of motion and constraints describe a {\bf multi-string}
configuration. That is, each branch of  $\tau$ as a function
of $T$ corresponds to  a different string.

This {\it multiple} number of strings arises as a consequence of the
string dynamics in curved spacetimes, that is,
from  the {\it coupling} of the string  with the spacetime
geometry.  Notice that here we have just {\it free} string equations
of motion in curved spacetime. That is, interactions between the
strings themselves, like splitting and merging,  are not considered.
We find that the geometry determines the simultaneous existence
of several strings. They do not interact directly between them since
they do not intersect. All the interaction is through the spacetime
geometry. Notice that such phenomenom does not appear in $D = 2$,
[eq.(\ref{enrl})] where time is a monotonic and periodic function of
$\tau$. This solution describes only one string in one period :
$0 < \tau \leq \pi/n $. For others periods we get identical
copies of the same string. This is not the case of the 2+1-dimensional
solutions displayed in figs.5-11. They describe five or three different
strings. Five is the generic number of strings in our dressed
solutions. This value five can be related to the fact that we
are dressing a one-string solution ($ q_{(0)} $) with {\it four}
poles. Each pole adds here an unstable string.

Figs. 7 and 8 show the function $ \tau = \tau(\sigma,
T)$ as a function of $\sigma$ -the
different values of $T$ are indicated in the pictures- for the
above solutions (i.e. polarization vectors $|x^0>$ and windings
$(n,m)$ the same as above). The function $\tau = \tau(\sigma,T)$
being periodic in $\sigma$ , it is plotted only for one period
($2\pi$) of  $\sigma$.
We see that {\it in addition} to the period $2\pi$, {\it another} period
in $\sigma$ appears which depends on $\tau$. $\tau = \tau(\sigma,
T)$ is a sinusoidal type function. It is more convoluted for small values
of $|\tau|$ in the neighbourhood of $\tau = 0$ where several maxima
and minima appear. As soon as the neighbourhood of $\tau = 0$
is left, $\tau = \tau(\sigma, T)$ becomes
very fast a regular sinusoidal-type
function of $\sigma$ with a fixed period much smaller than $2\pi$.
(In all solutions studied here, $\tau = \tau(\sigma,T)$ reaches
this asymptotic form for  $\tau \sim 5 $).
The meaning of these small and large $\tau$ behaviours will become  more
clear in connection with the evolution of the spatial  coordinates and
shape of the string. The small $ \tau $ behaviors are
connected  with the different (and complicated) ways in which the string
winds at the begining of its evolution, while the $ \tau \to
\infty$ uniform behavior is connected with the asymptotic configuration
which is 'frozen' in comoving coordinates. The large $\tau$
behaviour turns to be $\tau$-independent in comoving coordinates.

Let us analyze now the spatial coordinates $X^1(\sigma,\tau)$ and
$X^2(\sigma,\tau)$ of this solution.
Figs. 9-10 show the time evolution of the {\bf three} or
{\bf five} strings {\bf simultaneously} described by this solution.
In order to describe the real {\it physical} evolution, we eliminated
$\tau = \tau(\sigma, T) $ from the solution and expressed
 $X^1(\sigma,\tau) = X^1(\sigma, T)$ and
$X^2(\sigma,\tau) = X^2(\sigma,T)$
in terms of $T$. This was done numerically.
Figs. 10 show the comoving coordinates $(X^1, X^2)$ for
different times $H T$.
We see that for the fifth string, $(X^1, X^2)$ collapse
precisely as the inverse of the expansion factor $e^{-H T}$ ,
while the other four strings keep  $(X^1, X^2)$ constant in time
(in Fig. 9, it is the third string that collapses). That is, the
first string keep its proper size constant while the proper size of
the other four strings expand like  $e^{ H T}$.
These exact solutions display
remarquably the string behaviour found asymptotically and
approximately in refs.\cite{gsv}.
In summary, when $(X^1, X^2)$ are smaller or equal than $1/H$ (the
horizon radius), they contract to a point keeping the proper
amplitudes $(e^{ H T}\,X^1, e^{ H T}\,X^2)$ and proper size constant.
When $(X^1, X^2)$ are larger than $1/H$, they become very fast
constant in time, the proper size expanding with the universe
itself as   $e^{ H T}$ (string unstability).

In terms of the sinh-Gordon description
[see eqs.(\ref{tama})-(\ref{senh}) and fig. 1] ,
this means that for strings outside
the horizon, the sinh-Gordon function $\alpha(\sigma,\tau)$
for most of the history is the same as the cosmic time $T$ up to
a function of $\sigma$. We find combining eq.(5.14) in ref.\cite{prd}
with eq.(\ref{traf}):
\begin{equation}
\alpha(\sigma,\tau) \buildrel{T >> {1\over H} }\over =
2 H\, T(\sigma,\tau) + \log\left\{2 H^2 \left[ (A^{1}(\sigma)')^2 +
  (A^{2}(\sigma)')^2 \right] \right\} + O(e^{-2HT}).
\label{alfaa}
\end{equation}
Here $A^1(\sigma)$ and  $A^2(\sigma)$ are the $X^1$ and $X^2$
coordinates outside the horizon. For $T \to \infty$ the string is at
the absolute {\it minimun} $\alpha = + \infty $
of the sinh-Gordon potential
and possess an infinite size.

The string inside the horizon corresponds to the {\it maximun} of
the potential, $\alpha = 0$. This is the stable string with contracting
coordinates $(X^1, X^2)$ and {\it constant} proper size, appearing in all
the multi-string solutions found here. The value $ \alpha = 0$ is the
only in which the string can stay without being pushed down by the
potential to $\pm \infty$. This also explains why only one stable
string appears : it is not possible to put more than one string at
the maximun of the potential without falling down.
The starting zero soliton solution $\alpha = 0$ we have dressed
is a particular and very simple stable string.

For degenerate choices of $|x^0\rangle$, the number of strings reduces
to three [see fig. 6]. For large positive $T$ two of the strings
(strings 1 and 2) are of the unstable type and one (string 3) is of
stable type. In addition, strings 1 and 2 become identical in the
infinite $T$ limit. In fig. 9, we plot this solution for negative $T$.
We see that string 2 is stable for $T \to -\infty $ (it has constant
invariant size in such limit) whereas the invariant sizes of
strings 1 and 3 collapse
in this limit. In addition, there is an intermediate regime for
$|T| \leq 3 $ where the comoving size of the strings decreases
by a factor of about 10.

The features above described are {\it generically} exhibited by our
one-soliton multistring solutions independently of the
particular initial state of the string. (Fixed by the values $|x^0> $ and
$(n,m)$).

It is interesting to see how the shape of the string becomes more
symmetric for special values of $|x^0>$. For instance, a rosette
shape or a circle with many festoons are clearly shown by figs. 9-11.
They correspond to  $|x^0> = (1, -1, i , 1) $
with $ n = 4, 6 $, respectively.
These particularly symmetric vectors  $|x^0>$ yield also
degenerate solutions, in the sense that they contain only
three different strings instead of five,
as it happens in the generic case.

We also see that for the symmetric inital conditions for the
string state $|x^0>$, the function $\tau = \tau(\sigma, T) $
becomes a perfectly
symmetric periodic sinusoidal inside the period $2 \pi$, for
{\it all} values of $ \tau $ (including small $ |\tau| $ ), and
the additional very small period is practically the same for all
$|\tau|$ .

The number of string windings and festoons is related to the frequencies
in eq.(\ref{pitg}) and express in terms of $(n,m)$.
The $\sigma$ dependence is characterized by the frequencies [see
eq.(\ref{pitg}) ]
$$ \Omega_1 = \frac{2mn}{m^2+n^2} ,
\Omega_2 = \frac{m^2-n^2}{m^2+n^2}$$
and the basic frequency
$$\Omega_0 = { 1 \over {n^2 + m^2}} {\rm ~for~} n^2 + m^2 {\rm ~odd~},
\Omega_0 = { 2 \over {n^2 + m^2}} {\rm ~for~} n^2 + m^2 {\rm ~even~}.$$
For $ m = 1 $, the highest available frequency is the sum
$\Omega_1 + \Omega_2 = {{n^2 + 2 n - 1 } \over {n^2 + 1}}$.
This highest frequency determines the small period in
$ \tau = \tau(\sigma,
T)$ as a function of $\sigma$ for fixed large  $\tau$. That is,
$$2\pi {{n^2 + 1}\over{n^2 + 2n -1}}.$$
 In addition,
$${{\Omega_1 + \Omega_2 }\over  {\Omega_0}} = n^2 + 2n -1 {\rm ~for ~odd~} n,
{{\Omega_1 + \Omega_2 }\over  {\Omega_0}} =
  (n^2 + 2n -1)/2  {\rm ~for~ even~} n$$
gives the number of festoons
in the strings at a given $T$ (see figs. 7-8 and 9-11).

Strings propagating in de Sitter spacetime enjoy as conserved
quantities those associated with the O(3,1) rotations on the
hyperboloid (\ref{l1}). They can be written as
$$
L = \int_0^{2\pi} d\sigma \left(\; |q \rangle \;\langle{\dot q}|
 - |{\dot q}\rangle \;\langle q |
\; \right)J ~ = ~ {1 \over 2}\int_0^{2\pi} d\sigma ~( U + V)
$$
In order to compute $L$ it is convenient to relate $U$ and
$V$ with $U_{(0)}$ and $V_{(0)}$ using
eqs.(\ref{phixi1})-(\ref{phieta1})
   and the
asymptotic behaviour of $\Phi(\lambda) $ for
$ \lambda \to \infty $
[see eq.(\ref{phi})].
$$
\Phi(\lambda) = 1 + {{C(\eta,\xi)}\over {\lambda}} +
O({1 \over {\lambda^2}})
$$
where $C(\eta,\xi)$ is a matrix. We then find
$$
U + V = U_{(0)} + V_{(0)} + 2\; C(\eta,\xi)_{\sigma}
$$
Since $C(\eta,\xi)$ is a periodic function of $\sigma$,
$$
L = L_{(0)}
$$
for all solutions considered here. We recall \cite{dms} that
only $L_{01}$ does not vanish for $q_{(0)}$, taking the value
[see eq.(\ref{escal})] :
$$
L_{10} = - L_{01} = (n^2 + m^2) \pi
$$

\newpage

{\bf Figure Captions}:

\bigskip

{\bf Figure 1:} Effective potential corresponding
to the sinh-Gordon model.

\medskip
{\bf Figure 2:} Function $HT(\tau,\sigma)$, for fixed
$\sigma=.41$, for the $n=1$ string solution in $1+1$-de Sitter
spacetime.
\medskip

{\bf Figure 3:} Evolution of the string in $(\sigma, \tau)$ variables.
The three projections $(q^1,q^2), (q^1,q^3)$ and $(q^2,q^3)$ are shown
for $n=2, m=1$ and $\tau = 0.1$ and $2$ for $|x^0> = (1,-1,.1,.1i)$.

\medskip

{\bf Figure 4:} Same as fig.3  for
$n=5, \tau=0.5, 3.85$ and $|x^0> =(1,-1,i,1)$

\medskip

{\bf Figure 5:} Plot of the function $H T(\tau)$, for two
values of $\sigma$, for $n = 4, |x^0> = (1+i,.6+.4i,.3+.5i,.77+.79i)$.
The function $\tau(T)$ is multivalued,
revealing the presence of five strings.

\medskip

{\bf Figure 6:} Same as fig.5, for $n = 4, |x^0> = (1,-1,i,1). $
Because of a degeneracy, there are now only three strings.

\medskip

{\bf Figure 7:} $\tau = \tau(\sigma,T)$ for fixed $T$
for $n = 4, |x^0>=(1,-1,i,1)$. Three values of HT are
displayed, corresponding to HT=0 (full line), 1
(dots), and 2 (dashed line). For each HT, three curves are
plotted, which correspond to the three strings. They are ordered with
$\tau$ increasing.

\medskip

{\bf Figure 8:} Same as fig. 7 for
$n = 4, x^0>=(1+i,.6+.4i,.3+.5i,.77+.79i).$
a) The five curves corresponding to the five strings at HT=2.
b) The five curves for three values of HT: HT=0 (full line), 1
(dots), and 2 (dashed line).
\medskip

{\bf Figure 9:} Evolution of the three strings,
for $n = 4, |x^0> = (1,-1,i,1)$.
The comoving size of string (1) stays constant
for $H T <-3$, then decreases
around $H T = 0$, and stays constant again after $HT = 1$.
The invariant size of string (2) is constant for negative $HT$,
then grows as the expansion factor  for $HT > 1$,
and becomes identical to string (1).
The string (3) has a constant comoving size for $ HT < -3$,
then  collapses as $e^{-HT}$ for positive $HT$.

\medskip

{\bf Figure 10:} Evolution of three of the five strings for
$ n = 4, |x^0> = (1+i,.6+.4i,.3+.5i,.77+.79i)$.

\medskip

{\bf Figure 11:} Evolution of the three strings
for the degenerate case $n=6, |x^0> = (1,-1,i,1)$.


\newpage

\begin{thebibliography}{10}

\bibitem{dvs87}H J de Vega and N S\'anchez, Phys. Lett.{\bf B 197}, 320
(1987).
\bibitem{eri} See for a review the contributions by H J de Vega and
by N S\'anchez in  ``String Quantum Gravity and the Physics at
the Planck energy Scale'', Proceedings of the Erice Workshop
held in June 1992. Edited by N. S\'anchez,
World Scientific, 1993.
\bibitem{opg}H. J. de Vega and N. S\'anchez,
Phys. Rev. {\bf D 45} , 2783 (1992).

 H. J. de Vega, M. Ram\'on Medrano and N. S\'anchez,   LPTHE  Paris
  preprint  92-13.  \hfill \break
To appear in Classical and Quantum Gravity.
\bibitem{sw}H. J. de Vega and N. S\'anchez,
Nucl. Phys. {\bf B 317}, 706 (1989) .

 D. Amati and K. Klim${\rm \check{c}}$ik,
Phys. Lett. {\bf B 210} , 92 (1988)  ,

 M. Costa and H. J. de Vega, Ann. Phys. {\bf 211}, 223 and 235 (1991).

C. Loust\'o and N. S\'anchez, Phys. Rev. {\bf D46}, 4520 (1992).
\bibitem{agn}H. J. de Vega and N. S\'anchez,
Nucl. Phys. {\bf B309}, 552 and 577 (1988).

C. Loust\'o and N. S\'anchez, Phys. Rev. {\bf D47}, 4498 (1993).

\bibitem{con}H. J. de Vega and N. S\'anchez,
Phys. Rev. {\bf D 42} , 3969 (1990) and

 H. J. de Vega, M. Ram\'on Medrano and N. S\'anchez,
Nucl. Phys. {\bf B 374}, 405 (1992).
\bibitem{gsv}N. S\'anchez and G. Veneziano,
Nucl. Phys. {\bf B 333}, 253 (1990),

 M. Gasperini, N.S\'anchez and G. Veneziano,

 Int. J. of Mod. Phys. {\bf A 6}, 3853 (1991) and
Nucl. Phys. {\bf B 364}, 365 (1991).
\bibitem{zm1}V E Zakharov and A V Mikhailov, JETP, {\bf 75}, 1953 (1978).
\bibitem{twbk}See for a review, T.W.B. Kibble, Erice Lectures at the
Chalonge School in Astrofundamental Physics,
N. S\'anchez editor, World Scientific, 1992.

\bibitem{vil} A. Vilenkin, Phys. Rev. {\bf D 24},  2082 (1981),
 Phys. Rep.{\bf 121}, 263 (1985) .

N. Turok and P. Bhattacharjee, Phys. Rev. {\bf D 29}, 1557 (1984).

\bibitem{tuba} N. Turok, Phys. Rev. Lett. {\bf 60}, 549 (1988).
\hfill \break
J. D. Barrow, Nucl. Phys. {\bf B310}, 743 (1988).
\bibitem{dv78} H J de Vega,  Phys. Lett. {\bf B 87}, 233 (1979).
\bibitem{zm2}A V Mikhailov, Physica  {\bf D 3}, 73 (1981).
\bibitem{prd}H J de Vega and N S\'anchez,
Phys. Rev. {\bf D47}, 3394 (1993).
\bibitem{dms}H J de Vega, A V Mikhailov and N S\'anchez,
LPTHE preprint 92/32 (hep-th/9209047), Teor. Mat. Fiz. {\bf 94}, 232 (1993).
 (see also \cite{eri}).
\bibitem{ani}H J de Vega and I L Egusquiza, LPTHE preprint 93-43
(hep-th/9309016).
\bibitem{sol} Solitons and the Inverse scattering
transformation, M J Ablowitz and H Segur, SIAM Philadelphia 1981.

V.E. Zakharov, S.V. Manakov, S.P. Novikov and L.P. Pitaevsky
"Soliton Theory; The Inverse Method", Nauka, Moscow, 1980
\end{thebibliography}
\end{document}